\documentstyle[prl,aps,psfig,preprint]{revtex}
\begin{document}
\title{Non-Markovian Configurational Diffusion and Reaction Coordinates for
Protein Folding}
\author{Steven S. Plotkin and Peter G. Wolynes}
\address{Department of Physics and School of Chemical Sciences, 
University of Illinois, Urbana, IL, 61801 }
\date{\today}
\maketitle
\baselineskip=.5cm 

\begin{abstract}
\baselineskip=.5cm 
The non-Markovian nature of polymer motions is accounted for in
folding kinetics, using 
frequency-dependent friction. Folding, like many other problems in the
physics of disordered systems, involves barrier crossing on a
correlated energy landscape.
A variational transition state theory (VTST) that reduces to the usual
Bryngelson-Wolynes Kramers approach when the non-Markovian aspects are
neglected is used to obtain the rate, without making any assumptions
regarding the size of the barrier, or the memory time of the friction.
The transformation to collective variables dependent on the dynamics
of the system allows the theory to address the
controversial issue of what are ``good'' reaction coordinates for folding.
\end{abstract}
\pacs{PACS numbers: 64.60.Cn, 02.50.Ey, 61.41.+e}


\def \zn{z_{\mbox{\tiny N}}}
\def \zbar{\overline{z}}
\def \Zbar{\overline{Z}}
\def \znq{z_{\mbox{\tiny NQ}}}
\def \en{\epsilon_{\mbox{\tiny N}}}
\def \dele{\delta \epsilon}
\def \delen{\delta \epsilon_{\mbox{\tiny N}}}
\def \kboltz{k_{\mbox{\tiny B}}}
\def \tf{T_{\mbox{\tiny F}}}
\def \tg{T_{\mbox{\tiny G}}}
\def \ta{T_{\mbox{\tiny A}}}
\def \tgo{T_{\mbox{\tiny G}}^o}
\def \tc{T_{\mbox{\tiny C}}}
\def \etaH{\eta_{\mbox{\tiny H}}}
\def \etaC{\eta_{\mbox{\tiny C}}}
\def \znc{z_{\mbox{\tiny N}_{\! c}}}
\def \znh{z_{\mbox{\tiny N}_{\! h}}}
\def \zh{z_{\mbox{\tiny N$-$NQ}}}
\def \NH{N_{\mbox{\tiny H}}}
\def \NC{N_{\mbox{\tiny C}}}
\def \etaHqz{\etaH\left( Q,\zbar \right)}
\def \etaHzb{\etaH\left( \zbar/\zn \right)}
\def \etaHE{\eta_{\mbox{\tiny H}}^{\mbox{\tiny E}}}
\def \etaHL{\eta_{\mbox{\tiny H}}^{\mbox{\tiny L}}}
\def \ebar{\overline{\epsilon}}
\def \qmg{Q_{\mbox{\tiny MG}}}
\def \zmg{\zbar_{\mbox{\tiny MG}}}
\def \qf{Q_{\mbox{\tiny F}}}
\def \qu{Q_{\mbox{\tiny U}}}
\def \zf{\zbar_{\mbox{\tiny F}}}
\def \qst{Q^{\star}}
\def \zst{\zbar^{\star}}
\def \delF{\Delta F^{\star}}
\def \gtrsim{ \leavevmode{\raisebox{-.5ex}{ $\stackrel{>}{\sim}$ } } }
\def \lessim{ \leavevmode{\raisebox{-.5ex}{ $\stackrel{<}{\sim}$ } } }
\def \Evect{\left( \epsilon ,\ebar , \delen , \tf \right)}
\def \slev{s_{\mbox{\tiny LEV}}}
\def \le{\ell_{\mbox{\tiny E}}}
\def \lc{\ell_{\mbox{\tiny C}}}
\def \lec{\ell_{\mbox{\tiny EC}}}
\def \delsq{\Delta E^2}
\def \tbar{\overline{\tau}}
\def \tauo{\tau_{\mbox{\scriptsize o}}}
\def \tFbar{\overline{\tau_{\mbox{\tiny F}}}}
\def \maxQz{\leavevmode{\raisebox{-1.0ex}{$\stackrel{\mbox{\scriptsize{max}}}
	{\mbox{\scriptsize{Q, $\zbar$}}} $}} }
\def \Sqz{S\left(Q,\zbar\right)}
\def \dE2{\Delta E^2\left(Q,\zbar\right)}
\def \dE{\Delta E\left(Q,\zbar\right)}

\def \wo{\omega_o}
\def \qdag{q^\dag}
\def \qddag{q^\ddag}
\def \Qddag{Q^\ddag}
\def \adag{\alpha^\dag}
\def \ldag{\lambda^{\ddag}}
\def \ldagsq{\lambda^{\ddag 2}}
\def \wdag{\omega^{\ddag}}
\def \wdagsq{\omega^{\ddag 2}}
\def \zL{z_{\mbox{\tiny L}}}
\def \zU{z_{\mbox{\tiny U}}}
\def \zM{z_{\mbox{\tiny M}}}
\def \e{\mbox{e}}
\def \Fhyp{F_{_{\!\!\!\!\!\!\!\! 2 \;\, 1}}}
\def \lambGH{\lambda^{\mbox{\tiny GH}}}
\def \lambKR{\lambda^{\mbox{\tiny KR}}}
\def \kKR{k^{\mbox{\tiny KR}}}
\def \kGH{k^{\mbox{\tiny GH}}}
\def \En{E_{\mbox{\tiny N}}}
\def \sQ{\sigma_{\mbox{\tiny Q}}}
\def \FQ{F_{\mbox{\tiny Q}}}
\def \SQ{S_{\mbox{\tiny Q}}}
\def \PQ{P_{\mbox{\tiny Q}}}
\def \AQ{A_{\mbox{\tiny Q}}}
\def \tQ{\tau_{\mbox{\tiny Q}}}
\def \tgh{t_{\mbox{\tiny GH}}}
\def \tkr{t_{\mbox{\tiny KR}}}
\def \pq{p_{\mbox{\tiny Q}}}
\def \pro{p_{\rho}}
\def \psig{p_{\sigma}}
\def \pxj{p_{x_{j}}}
\def \pyj{p_{y_{j}}}
\def \lambQ{\lambda_{\mbox{\tiny Q}}}
\def \gammaQ{\gamma_{\mbox{\tiny Q}}}
\def \Vb{V_{\beta}}

\renewcommand{\theequation}{\arabic{equation}}
\narrowtext

According to the energy landscape theory~\cite{BryngelsonJD89},
protein folding can be seen as a stochastic motion
of a few collective coordinates describing protein conformation,
on an average thermodynamic potential~\cite{PlotkinSS97}. 
To first approximation this motion is Brownian, and the
folding time can be computed from diffusive rate
theory~\cite{SocciND96:jcp,Thirumalai97}.  
A good quantitative comparison between the analytical energy landscape
theory and lattice simulations of smaller proteins
has been made~\cite{SocciND96:jcp}; 
The diffusive behavior of the reaction coordinate's motion is approximate;
in lattice simulations a fraction of the
trajectories have {\it ballistic}
crossings over the barrier, while others are quite
diffusive~\cite{OWLS}, suggesting a wide range of time scales for the
collective reaction coordinate.
Non-Markovian dynamics, expected on a rugged energy landscape,
will affect reaction rates when
the time to cross the top of the barrier is
comparable to the memory time of the fluctuating forces acting on
the collective coordinate.
Analysis of such situations for reactions in condensed
phases has led to a number of good 
approximations for rates~\cite{Hynes80,Onuchic88,Pollak90,Berezhkovskii92c}.
In our treatment we use variational transition-state theory
(VTST)~\cite{Pollak90,Berezhkovskii92c,Pollak91} to discuss how
non-Markovian dynamics of the chain affects folding, and to address the
question of what is the best reaction  
coordinate for folding.
We apply our results to the motion of a heteropolymeric protein
chain, but many of the same issues occur for kinetics in other
disordered systems, for example the nucleation of a crystal 
from a glassy liquid.
We first discuss the form of the effective frequency-dependent friction
$\widehat{\zeta}(\omega)$ for motion on a correlated energy
landscape, specifically for a heteropolymer.
Then we apply VTST to find corrections to the Kramers folding rate
due to memory effects in $\widehat{\zeta}(\omega)$, and
anharmonicities in the potential.

Protein conformational motion can
be mapped onto a generalized master equation, with 
escape rates from a statistical ensemble of configurations given in
terms of a waiting time distribution
$\PQ(\tau,\tilde{T})$~\cite{BryngelsonJD89}.  
The configurational states may be grouped together in strata with a
common value of their similarity to the 
native conformation, which is
an approximate reaction coordinate for folding. 
This is
often taken to be $Q$~\cite{SaliA94:nat,OWLS}, the fraction of native
contacts, but other choices are possible.
Finding a ``best'' reaction coordinate is currently of great
interest~\cite{OWLS,BoczkoEM95,DillKA97,Pande97pre}.
By projecting onto this coordinate, a diffusion equation for its
probability distribution is obtained with a
frequency-dependent diffusion coefficient~\cite{BryngelsonJD89}.  
Correspondingly, the coordinate's motion can be characterized by an
overdamped generalized Langevin equation (GLE): 
\(
-dF(Q)/dQ - \int_0^t d\tau \zeta_{\mbox{\tiny 
Q}}(t-\tau,T) \dot{Q}(\tau) + \xi(t)
= 0 \: ,
\)
with a frequency-dependent friction coefficient
$\widehat{\zeta}_{\mbox{\tiny Q}}(\omega,T) = \kboltz
T/\widehat{D}_{\mbox{\tiny Q}}(\omega)$ satisfying
$\langle \xi(t) \xi(t') \rangle = T
\zeta_{\mbox{\tiny Q}}(t-t',T)$.

The GLE implies that $Q$ responds linearly to
fluctuations in the other coordinates of the polymer chain apart from
the nonlinearity inherent in the thermodynamic potential for $Q$.
This should be a good approximation if many individual
configurational states of the polymer chain are sampled for each value
of $Q$, as expected above the glass
transition temperature of the stratum at $Q$.
$\widehat{\zeta}_{\mbox{\tiny Q}}(\omega,T)$
is given by averaging over $\PQ(\tau,T)$ as 
\(
\widehat{\zeta}_{\mbox{\tiny Q}}(\omega,T) = \lambQ
\langle \tau/(1+\omega \tau) \rangle /
\langle 1/(1+\omega \tau) \rangle \)
or 
\( 
\lambQ {\cal L}_t  \langle \mbox{e}^{-t/\tau}\rangle /
{\cal L}_t \frac{-d}{dt} \langle \mbox{e}^{-t/\tau}\rangle
\)
where ${\cal L}_t$ is the Laplace transform. 
The conformational motion distance scale is set
by $\lambQ \equiv 2 \kboltz T/(\Delta Q^2 \gammaQ)$, where
$\Delta Q$ is the step size  and $\gammaQ$ is the probability a
jump changes $Q$.
$\PQ(\tau,T)$ on a {\it correlated} energy landscape 
is obtained by first
defining a local progress coordinate $q$ as similarity to the
{\it given} trap state, so the typical rate of
escape~\cite{WangPlotJ97} involves the calculation of 
the free energy barrier $(\FQ(\qddag) - \FQ(1))$ for 
motion to states having native similarity near to $Q$,
assuming the elementary moves are sufficiently local.  
A bilinear approximation
to the entropic part of $\FQ(q)$ can be used~\cite{WangPlotJ97}, since
contacts formed at 
small $q$, i.e. for a more
weakly constrained polymer, cost more entropy than for a strongly
constrained one at high $q$.
The escape time $\propto
\exp(\Delta F^\ddag/T)$ depends on 
2 parameters: 1.) The reduced temperature
$\tilde{T} = T/T_G$, where $T_G = (\Delta E_Q^2/2 \SQ)^{1/2}$ is the glass
temperature for the $Q$ stratum. $\SQ$ is the
configurational entropy at $Q$ while $\Delta E_Q^2 =
(1-Q)(\Delta E_M^2 + Q \Delta E_N^2)$ is the
energetic variance of the states in terms of the variance of native
(N) and non-native (M) contacts, and 
2.) The reduced energy of the trapped state $\tilde{E}=E/E_{GS}$,
where $E_{GS} = -(2 \SQ \Delta E_Q^2)^{1/2}$ is the ground state
energy of the ensemble of states at $Q$.
The escape time from a trap with energy $\tilde{E}$, $\tau_Q(\tilde{E},\tilde{T})$, is 
given by $ \tau_o \exp [ \SQ (1-\qdag)
( 2 \tilde{E}/\tilde{T} - 1/\adag - 1/\tilde{T}^2) \theta_r]$.
Here  $\qdag$ is the location of the 
barrier peak, and $\adag = 
(1-\qdag) \SQ /S^\dag$, where $S^\dag$ is the fraction of
$\SQ$ at the barrier peak (for the $64$-mer $\qdag \cong 0.3$ and
$\adag \cong 1.6$, see fig. 4-9 of~\cite{WangPlotJ97}). 
For states with $\tilde{E}<\tilde{E}_A(\tilde{T})$ or when 
$\tilde{T}>\tilde{T}_A(\tilde{E})$ (determined by setting the barrier
 to zero), escape becomes downhill with short life-times of roughly the
Rouse-Zimm time(s) $\simeq \tau_o$, hence
$ \theta_r = 
\theta ( \tilde{E}-\tilde{E}_A(\tilde{T}))\theta
(\tilde{T}_A(\tilde{E})-\tilde{T})$ in the exponent.
At $\tilde{T}=1$ where
the system is frozen into one of a few ground states ($\tilde{E}=1$), 
and the typical escape time is
$\tau_Q(1,1) = \tau_o \exp [ \SQ (1-\qdag)
(1- 1/\adag ) ] \cong \tau_o \exp ( 0.27 \,
\SQ^{(64)} )$.
Correlations lower the free energy barrier to correspond to roughly
$1/4$ the total configurational entropy~\cite{WangPlotJ97}, as opposed
to the full entropy as in uncorrelated landscapes.
The distribution of occupied state energies $\tilde{E}$ at temperature
$\tilde{T}$ is a Boltzmann weighted gaussian:
$\PQ(\tilde{E},\tilde{T}) \sim \exp [-\SQ(
\tilde{E}-1/\tilde{T})^2]$. 
Reflecting this, the distribution of escape times is
easily calculated as  
$\PQ(\tau,\tilde{T}) = \int_{-1}^{1} d\tilde{E} \,
\PQ(\tilde{E},\tilde{T})  
\delta[\tau-\tau_Q(\tilde{E},\tilde{T})] = \PQ^{\mbox{\tiny B-L}} +
\PQ^{\mbox{\tiny B}}$.  Apart from
barrier-less escapes with $\PQ^{\mbox{\tiny B-L}} (\tau,\tilde{T}) = \Delta(\tilde{T})
\delta(\tau-\tau_o)$, this yields essentially a log-normal distribution:
\begin{eqnarray}
\PQ^{\mbox{\tiny B}} (\tau,\tilde{T}) &=& 
\theta_{\mbox{\tiny B}} \frac{\tilde{T} A}{2
S\left(1-\qdag\right)}\frac{1}{\tau}
\mbox{e}^{-\frac{S}{4}\left[\frac{\tilde{T} \ln 
\left(\tau/\tau_o\right)} {S\left(1-\qdag\right)}
-\frac{1}{\tilde{T}}+\frac{\tilde{T}}{\adag} \right]^2}
\nonumber 
\end{eqnarray}
where $\theta_{\mbox{\tiny B}} = \theta(\tilde{T}_A(1)-\tilde{T}) 
 \theta(\tau-\tau_o)\theta(\tQ(1,\tilde{T}) -\tau)$,
$S \equiv \SQ$, 
and $A = \AQ(\tilde{T})$ is the normalization
constant of $\PQ(\tilde{E},\tilde{T})$, while
$\Delta = \Delta(\tilde{T})$ is the weight for fast escapes from states without
barriers. 

The above analysis applies at temperatures higher than the
thermodynamic glass transition
temperature $\tg$ (but below the activation temperature $\ta$, where
escape from all occupied states is barrier-less).
At or below $\tg$ the
temperature independent distribution of state energies becomes
$P(E) \sim \exp (E/T_G)$. 
Arrhenius-like escape from states yields the distribution of escape times
$P(E(\tau))/\left|\frac{d\tau}{dE}\right| \sim
\tau^{-(1+T/T_G)}$, giving stretched
exponential relaxations~\cite{Shlesinger84,Bouchaud95}.
In this regime the assumed linearity of $Q$ dynamics is questionable however.

Evaluating $\widehat{D}_{\mbox{\tiny Q}}(\omega,\tilde{T})$ or
$\widehat{\zeta}_{\mbox{\tiny Q}}(\omega,\tilde{T})$ above $\tg$,
note that $ \langle \mbox{e}^{-t/\tau}\rangle \sim$
$\int_{\zL}^{\zU} dz \, \mbox{e}^{-S z^2} \exp [-b
\frac{t}{\tau_o} \mbox{e}^{-c_2 S z}] $
(where $ c_2  = 2(1-\qdag)/\tilde{T} $, and $ b =$
$\exp[S(1-\qdag) (1/\adag - 1/\tilde{T}^2 )]$)
is reminiscent of the after-effect function
$f(bt/\tau_o, c_2/\sqrt{S})$ used for
non-exponential decay in glasses~\cite{Dimarzio86}. 
But for the mesoscopic systems relevant to folding ($N \lessim
100$), a better approximation is to linearize the gaussian term 
on the range $(\zL,\zU) = (\tilde{T}/2\adag -1/2\tilde{T},1 - 1/\tilde{T})$, 
yielding a closed form expression for the friction
\begin{equation}
\widehat{\zeta}_{\mbox{\tiny Q}}\left(\omega,\tilde{T}\right) = \lambQ 
\frac{\frac{\Delta}{1+\omega} +
\frac{F}{1-c} \left[ f^{c-1} \; \Fhyp
\left(1,1-c,2-c,-\omega/f\right) - \Fhyp \left(
1,1-c,2-c,-\omega\right) \right]}
{\frac{\Delta}{1+\omega} + \frac{F}{c} \left[ \; \Fhyp
\left(1,-c,1-c,-\omega\right) - f^c \; \Fhyp
\left(1,-c,1-c,-\omega/f\right)\right] }
\label{eq:zwx}
\end{equation}
where $\Fhyp$ is the hypergeometric function, $\omega$ is in
units of $1/\tau_o$, 
$ F = A \e^{S \zM^2}/c_2 b^c S $, $ c = 2 \zM/c_2 $, 
$ f = \tau_o/\tQ(1,\tilde{T}) < 1 $, and 
$ \zM = (\zU + \zL)/2$.
The corresponding diffusion coefficient $\widehat{D}_{\mbox{\tiny
Q}}(\omega,\tilde{T}) = \kboltz T/\widehat{\zeta}_{\mbox{\tiny
Q}}(\omega,\tilde{T})$ is plotted in fig. 1.

Non-Markovian rate theory~\cite{Pollak90} proceeds by
recognizing that the GLE is equivalent 
to a particle bilinearly coupled to a bath of
oscillators~\cite{Zwanzig73} with an effective Hamiltonian
$
{\cal H} = \pq^2/2 + F(Q) + \frac{1}{2} \sum_j  
[\pxj^2 + ( \omega_j x_j - c_j Q/\omega_j)^2 ]
$
with $\zeta(t) = \sum_j (c_j^2/ \omega_j^2) \cos(\omega_j t)$.
These additional collective modes describe the dynamics of $\dot{Q}$
fluctuations within linear response.
Rather than dealing with non-Markovian dynamics of $Q$ we can 
study the dynamics of this equivalent many-dimensional system without
memory.  When a 
single barrier exists in $F(Q)$ it makes sense to use
multidimensional transition state theory. 
If the barrier is large and its parabolic part dominates, ${\cal
H}$ is quadratic and may be diagonalized by a normal mode
transformation, which singles out as a reaction coordinate an
unstable mode $\rho$ with 
imaginary frequency $i \lambda^{\ddag}$ given by the solution of 
\(
\ldag \widehat{\zeta}_{\mbox{\tiny $\Qddag$}}(\ldag,\tilde{T}) = m
\omega^{\ddag 2}  = \left|\partial^2 F(Q) / \partial Q^2
\right|_{\Qddag} 
\).
This frequency is identical to the (overdamped) Grote-Hynes reactive
frequency~\cite{Hynes80}.
The effect of friction is to leave the
barrier height unchanged, but to rotate the reaction coordinate to a 
different direction in configuration space.
When the friction has no memory and the $Q$ motion is purely diffusion,
the reactive frequency is that of an overdamped inverted harmonic
oscillator corresponding to the Kramers prefactor in the rate.
For a general $F(Q)$ we can separate the quadratic part from the
anharmonic part of $F$. Then in the normal coordinates at the saddle
point $
{\cal H} = \frac{1}{2} [\pro^2 - \lambda^{\ddag 2} \rho^2 +
\sum_j (\pyj^2 + \lambda_j^2 y_j^2 ) ] + F_1(
u_{oo} \rho + \sum_j u_{jo} y_j ) $.
The coefficients $u_{jo}$ are elements of the orthogonal normal-mode
transformation such that $q \equiv Q-\Qddag = u_{oo} \rho + \sum_j
u_{jo} y_j$ 
( $u_{oo}$ is given in terms of the friction
kernel as $u_{oo}^2 = 1/[\widehat{\zeta}_{\mbox{\tiny
$\Qddag$}}(\ldag,\tilde{T})/\ldag +  \partial
\widehat{\zeta}_{\mbox{\tiny $\Qddag$}}(s,\tilde{T})/ \left. \partial s
\right|_{s=\ldag} ]$).
In addition to the Grote-Hynes coordinate $\rho$ one can define a
residual collective bath coordinate  $\sigma \equiv (1-u_{oo})^{-1/2}
\sum_j u_{jo} y_j \equiv (1/u_1) \sum_j u_{jo} y_j$, which appears in
the anharmonic part of the potential.
The effects of dynamic friction, reflected by re-crossings in the $Q$
coordinate, are 
in fact accounted for by ballistic motion across a new {\it
thermodynamically} determined dividing surface involving a 2-D
potential in $(\rho,\sigma)$~\cite{Pollak90}. 
The equilibrium flux across any dividing surface written in these new
coordinates is 
\begin{eqnarray}
\Gamma &=& \frac{\int d\pro \, d\rho \, d\psig \, d\sigma \:
\delta\left(f\right) 
\left(\nabla f \cdot {\bf p}\right) \theta\left(\nabla f \cdot {\bf
p}\right) \mbox{e}^{-\beta  {\cal H}^\ddag } }
{\int d\pro \, d\rho \, d\psig \, d\sigma \: \mbox{e}^{-\beta {\cal
H}^\ddag}  }	\nonumber
\end{eqnarray}
Here $f = \rho - g(\sigma)$ serves as a new progress coordinate.
$f=0$ determines a dividing surface between reactants and products,
the factor $\delta(f)$ localizes the integration to that
surface, while $\nabla f \cdot {\bf p} = \pro \partial f/\partial \rho +
\psig \partial f/\partial \sigma$ 
is the flux density across the surface.
The reduced Hamiltonian depends only on two coordinates $\rho$ and $\sigma$:
${\cal H}^\ddag = \frac{1}{2}
[\pro^2 +\psig^2 - \lambda^{\ddag 2} \rho^2 + \Omega^2
\sigma^2] + F_1( u_{oo} \rho + u_1 \sigma )$,
with a collective bath frequency $\Omega^2 =
u_1^2/(u_{oo}^2/\ldagsq - 1/\wdagsq)$.
Carrying out the integrations, the rate can be written as a correction
to the Grote-Hynes rate: $\Gamma = P (\ldag/\wdag)
\Gamma_0$. $\Gamma_0 = (\omega_o/2 \pi)
\exp(-\beta F^{\ddag})$ is the TST rate, while $\ldag/\wdag$ is
the G-H transmission factor. The additional correction to the
G-H prefactor is given by a quadrature:
\begin{eqnarray}
P &=& \left(\frac{\beta \Omega^2}{2 \pi}\right)^{1/2}
\int_{-\infty}^{\infty} d\sigma \, \sqrt{1+
\left(\frac{dg\left(\sigma\right)}{d\sigma}\right)^2} \,
\exp\left(-\beta E\left[g\right]\right) \: . \nonumber
\end{eqnarray}
Here $E[g] = \frac{1}{2}(\Omega^2 \sigma^2 - \ldagsq
g(\sigma)^2 ) + F_1(u_{oo}
g(\sigma)  + u_1 \sigma )$.
If $F_1$ vanishes, the potential is parabolic, and variational
minimization of the rate yields $f=\rho$ as the progress coordinate
and $q = u_1 \sigma \cong \sigma$ as the ideal dividing surface. The
transition state position is coupled strongly to the bath mode, but the
correction gives $P=1$, reproducing the G-H rate $\Gamma
= (\ldag/\wdag) \Gamma_0$.
For systems with large barriers $\gtrsim 10 \kboltz T$, the parabolic
approximation discussed above is highly accurate but we can find
corrections to it by finding a more general {\it planar} dividing surface 
$f= u_{oo} \rho + \sum_j u_{oj} y_j - \rho_o = 0$ (where $\rho_o$ is the
distance of the dividing surface from the barrier peak) oriented such
that the TST rate is minimized. Using the variational framework developed
in~\cite{Berezhkovskii92c} along with the correlated landscape
theory of $\widehat{\zeta}_{\mbox{\tiny Q}}(\omega,\tilde{T})$, 
we find corrections to the GH rate at conditions of
equilibrium between the folded and molten globule states (at
$\tf$), see fig. 2. 
The optimal barrier location $\rho_o$ (see inset A) is little different
from the naive choice of the free energy maximum, for the
$64$-mer. 
The rate enhancement over the Kramers result has a peak at
intermediate $\tf/\tg$ due to the larger 
relative frequency dispersion of the diffusion coefficient at
intermediate temperatures. 
For systems this large, the planar dividing
surface assumption is accurate to within  $5 \%$.

For a $27$-mer imitating a small protein at $\tf \cong 1.6 \tg$, 
we can use the simulated
auto-correlation function of $Q$, $c(t)$ (see fig. 9
of~\cite{SocciND96:jcp}), 
to determine $\widehat{\zeta}( \omega)$:
$\widehat{\zeta}( \omega) =
F''(\Qddag)/(1/\widehat{c}(\omega) -
\omega)$. Here $\widehat{c}(\omega) = {\cal L}_t
c(t)$. For the $27$-mer at $\tf$, the potential is 
very anharmonic. A planar dividing
surface is no longer optimal. For such low barriers, minimizing the
flux through the TST surface can be carried out using the calculus of
variations as done by Miller~\cite{MillerW74} and Pollak~\cite{Pollak91}.  
Finding the stationary point of the rate for arbitrary
variations of the dividing surface functional $\delta g(\sigma)$ yields a
differential equation for $g(\sigma)$:
\(
g''/(1+g'^2) = \beta (g' \partial
E[g]/\partial \sigma - \partial E[g]/\partial
g)
\).
Treating $g$ and $\sigma$ as parametrized variables in terms of an
independent parameter $t$ (such that $g' = \dot{g}/\dot{\sigma}$, and
setting the first integral $\frac{1}{2}(\dot{g}^2 + 
\dot{\sigma}^2) \equiv E_o - \Vb$)
recasts this variational equation into Hamilton's
equations of motion for $g$ and $\sigma$ on an effective temperature
dependent potential $\Vb = -(1/2 \beta) \exp(-2\beta
E[g])$ at total energy $\frac{1}{2}(p_g^2 +
p_\sigma^2) + \Vb = 0$.
The optimal dividing surface is a classical periodic trajectory on $\Vb$
with infinite period, that divides the
$(\rho, \sigma)$ space into reactants and products.
The correction $P$ is given in
terms of the action along this optimal trajectory: 
\(
P = ( \beta^2\Omega^2/2 \pi)^{1/2} \int ds \, \sqrt{-2 \Vb}
\)
where $ds$ is arc-length along the trajectory.
The optimal dividing surface for the 27-mer is plotted on the potential
$E(\rho,\sigma)$ in fig. 3. The correction $P \cong 0.85$.
The rate is moderately reduced from the
G-H value of $1.57 \, \kKR$, giving $k = 1.33 \, \kKR$ for the
corrected rate, in
closer agreement with the naive Kramers value.
While the Kramers approximation made
in~\cite{SocciND96:jcp,Thirumalai97} is numerically 
quite accurate, the optimal dividing surface shows the
transition region  is quite spread out in the coordinate $Q$. 
By finding the optimal dividing surface, VTST seeks that coordinate which
behaves most ballistically.
Including the $Q$ dependence of $\widehat{\zeta}$ 
at higher nativeness may explain
some of the discrepancy between simulations and rate
calculations~\cite{SocciND96:jcp}.

The present analysis can easily be generalized to include ordering
along additional collective coordinates, as for example the total
density of contacts, which is often important if collapse is not
fast~\cite{Socci97}.  Potentially interesting
effects may arise in such scenarios due to anisotropic 
friction~\cite{Berezhkovskii92c}, and
are a topic of future work.
The methods presented here are general, and also apply to other
condensed matter systems with rugged landscapes, e.g. glasses and
clusters.
Nonlinear couplings between the system and bath may also be treated;
This may allow explicit treatments of the 
deep traps in the glassy regime.
\begin{figure}[htb]
\hspace{0cm} 
\psfig{file=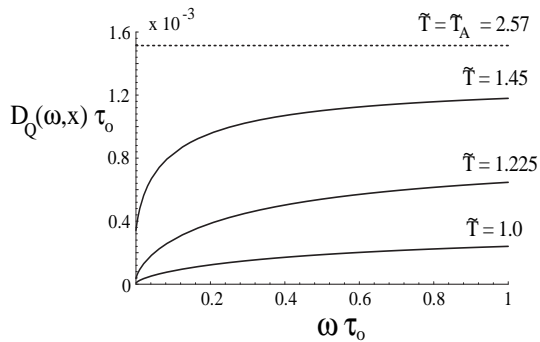,height=4.6cm,width=7cm,angle=-90}
\caption{
$\tau_o \widehat{D}_{\Qddag}
(\omega,\tilde{T})$ 
for $64$-mer near the transition state 
$Q^\ddag \approx 0.3 $, as
a function of frequency (in units of $\tau_o$), at
several temperatures between the thermodynamic glass
temperature and  kinetic (activated) glass temperature.
There is a rapid increase from the zero frequency value
$\widehat{D}_Q(0,\tilde{T})$ dependent on the typical escape time, to
a higher asymptotic value depending on how many of the states are untrapped
and have short lifetimes at that temperature. The dispersion in the
values of the diffusion is thus maximum at intermediate values of
temperature. The largest values of the diffusion constant are set by
$\sQ \equiv \kboltz T/\lambQ$, and at $\tilde{T}_A$,
$\widehat{D}_{\Qddag}(\omega,\tilde{T}_A) = \sQ$$= 
\Delta Q^2 \gammaQ/2 \tau_o \approx 0.0015/ \tau_o$ for the $64$-mer.}
\end{figure}
\newpage
\begin{figure}[htb]
\psfig{file=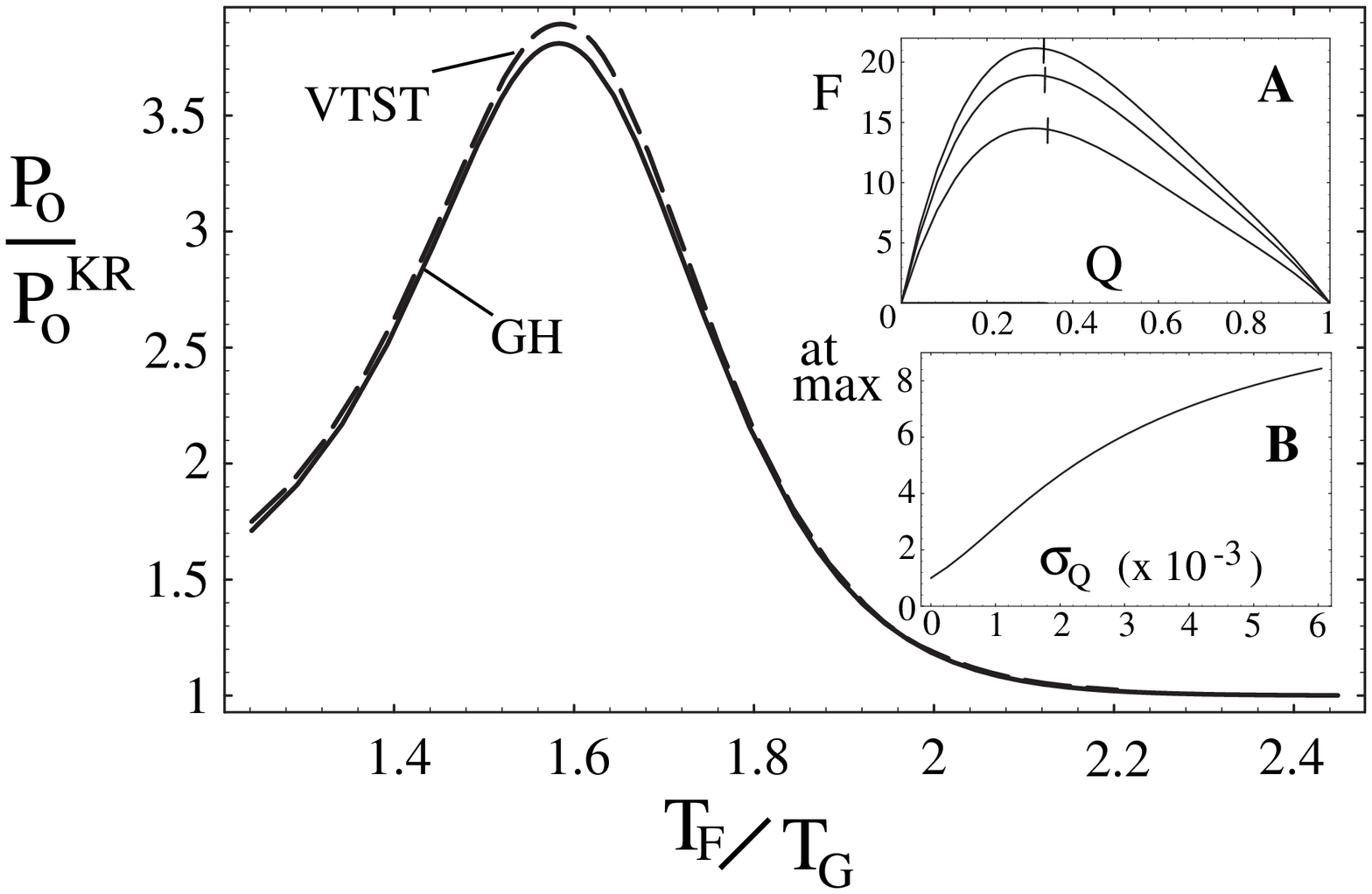,height=5.25cm,width=7cm,angle=-90}
\caption{}
\end{figure}
{\bf (solid line)} Ratio of the Grote-Hynes rate to the Kramers rate,
as a function of $\tf/\tg$. {\bf (dashed line)} VTST rate enhancement
including anharmonic effects of a finite size barrier. The effect here
is small, however larger effects are seen for shorter polymers (see text).  
The system is a $64$-mer at folding equilibrium.
The variance of interaction energies is varied, so that the
temperature ratio $\tf/\tg$ varies
($\tf/\tg = \sqrt{\eta} + \sqrt{\eta-1}$, with $\eta = \En^2/2 S_o
\Delta E_M^2$, where $\En$ is the native state energy, 
and $S_o$ is  total entropy). The rate closely follows the G-H result 
$\kGH/\kKR = \widehat{\zeta}_{\mbox{\tiny 
$\Qddag$}}( 0,\tilde{T}) /\widehat{\zeta}_{\mbox{\tiny
$\Qddag$}}( \lambGH,\tilde{T})$.
{\bf (Inset A)}  Thermodynamic potentials
vs. Q. For more rugged landscapes the barrier is flatter, and this
reduces the prefactor to the rate since there are more recrossings.
The $\tf/\tg$ values are $2.45$, $1.84$, and $1.24$ in order of
decreasing barrier size. The small vertical bars near the barrier peak
are where the VTST dividing surfaces cross the coordinate $Q$.
{\bf (Inset B)} Enhancement of the rate at the maximum value of $3.8$
at $T_{\mbox{\tiny F}}/T_{\mbox{\tiny G}} \approx 1.6$ by   
allowing the prefactor $\lambQ \equiv \kboltz T/\sQ$ in
$\widehat{D}(\lambGH, \tilde{T} )$ to vary ($\sigma_Q \approx
0.0015$ is the original value).

\newpage
\begin{figure}[htb]
\psfig{file=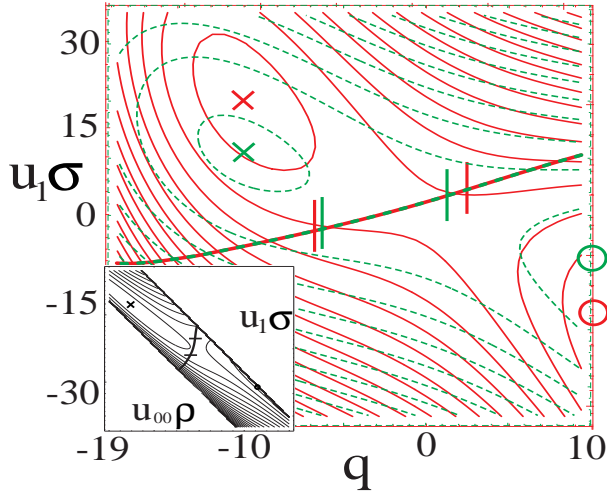,height=6.5cm,width=8cm,angle=0}
\caption{
(Color) Potential surface $E(q,u_1\sigma)$ (red) and in the normal
coordinates $(u_{oo}\rho,u_1 \sigma)$ (inset), along with the
variational dividing surface which minimizes the TST flux (heavy
line). Contours are drawn at intervals of about $2 \kboltz T$.
The potential in the Markovian case (dashed green), 
with the corresponding Kramers rate, is further skewed with respect to
the dividing surface, indicating paths in this case are even more
diffusive.{\bf X} marks the position of the molten globule minimum, and {\bf O}
the native minimum. The short vertical lines
(and horizontal lines in inset) bound a region of $\approx 70 \%$ of
the total flux across the dividing surface.
In $(q,u_1 \sigma)$ space there is flux over a wide
range of $q$ values, $\Delta q/(q_{native} - q_{mg}) \approx
0.44$, so that the transition state theory that reproduces the
multiple crossing physics in Kramers theory does not have
a well defined value of $q^\ddag$. 
However in $(u_{oo}\rho,u_1 \sigma)$ space the TST surface
tends towards orthogonality to the reaction coordinate $u_{oo} \rho$:
$\Delta \rho/(\rho_{native}-\rho_{mg}) \approx 0.04$, indicating
trajectories behave more ballistically along $\rho$. }
\end{figure}

\subsection{Acknowledgements}
We thank J. Onuchic, I. Rips, N. Socci, and J. Wang for helpful discussions.
This material is based upon work supported by the National Institutes of Health under 
award number PHS R01GM44557.

\vspace{-0.5cm}

\begin{thebibliography}{99}

\bibitem{BryngelsonJD89}
	J. Bryngelson and P. Wolynes, J Phys. Chem., {\bf 93}, 6902 (1989).
	
\bibitem{PlotkinSS97}
  S. Plotkin, J. Wang and P. Wolynes,
  J Chem. Phys. , {\bf 106}, 2932 (1997).

\bibitem{SocciND96:jcp}
  N. Socci, J. Onuchic and P. Wolynes, J Chem. Phys.,{\bf 104}, 5860 (1996).

\bibitem{Thirumalai97}
  D. Klimov and D. Thirumalai,
 Phys. Rev. Lett., {\bf 79}, 317 (1997).

\bibitem{OWLS}
  J. Onuchic, P. Wolynes, Z. Luthey-Schulten and N. Socci,
  Proc. Nat. Acad. Sci. USA, {\bf 92}, 3626 (1995).

\bibitem{Hynes80}
  R. Grote and J. Hynes, J Chem. Phys.,
  {\bf 73}, 2715 (1980).

\bibitem{Onuchic88}
  J. Onuchic and P. Wolynes, J Phys. Chem.,
  {\bf 92}, 6495 (1988).

\bibitem{Pollak90}
  E. Pollak, S. Tucker and B. Berne, Phys. Rev. Lett., 
  {\bf 65}, 1399 (1990).
		  
\bibitem{Berezhkovskii92c}
  A. Berezhkovskii, E. Pollak and V. Zitserman, J Chem. Phys., 
  {\bf 97}, 2422 (1992).
		  
\bibitem{Pollak91}
  E. Pollak, J Phys. Chem., {\bf 95}, 10235 (1991).

\bibitem{SaliA94:nat}
  A.{\u{S}}ali, E. Shakhnovich and M. Karplus, Nature, {\bf 369}, 248 (1994).

\bibitem{BoczkoEM95}
  E. Boczko and C. Brooks, Science, {\bf 269}, 393 (1995).

\bibitem{DillKA97}
  K. Dill and H. Chan, Nat. Struct. Biol., {\bf 4}, 10 (1997).

\bibitem{Pande97pre} R. Du, V. Pande, A. Grosberg, T. Tanaka and E. Shakhnovich,
  (preprint), (1997).

\bibitem{WangPlotJ97}
  J. Wang, S. Plotkin and P. Wolynes, J. Phys. I France, {\bf 7}, 395 (1997).

\bibitem{Shlesinger84}
  M. Shlesinger and E. Montroll, Proc. Nat. Acad. Sci. USA, {\bf 81}, 1280 (1984).

\bibitem{Bouchaud95}
  J-P. Bouchaud and D. Dean, J. Phys. I France, {\bf 5}, 265 (1995).

\bibitem{Dimarzio86}
  E. DiMarzio and I. Sanchez, in {\it Transport and Relaxation in Random Materials},
  ed. J. Klafter, R. Rubin and M. Shlesinger (World Scientific, Singapore, 1986).

\bibitem{Zwanzig73}
  R. Zwanzig, J Stat. Phys., {\bf 9}, 215 (1973).

\bibitem{MillerW74}
  W. Miller, J Chem. Phys., {\bf 61}, 1823 (1974).

\bibitem{Socci97}
  N.  Socci, J. Onuchic, P. Wolynes, Proteins: Struct. Funct. and Genetics, 
  (to appear)  (1997).		  

\end{thebibliography}

\end{document}